\documentclass[conference]{IEEEtran}
\IEEEoverridecommandlockouts

\def\BibTeX{{\rm B\kern-.05em{\sc i\kern-.025em b}\kern-.08em
    T\kern-.1667em\lower.7ex\hbox{E}\kern-.125emX}}
\usepackage{tikz}
\usepackage{pgfplots}
\usepackage[font=Large]{subfig}
\usepackage{booktabs} 
\usepackage{graphicx}
\usepackage{balance}
\usepackage{amsmath}
\usepackage{algorithm}
\usepackage{amsfonts}
\usepackage[noend]{algpseudocode}
\definecolor{darkgreen}{rgb}{0.125,0.5,0.169}

\newtheorem{definition}{Definition}

\newcommand{\spara}[1]{\smallskip\noindent{\bf{#1}}}

\def\ed{\sc{IPM}}
\def\baseNode{Competitor}

\begin{document}
\title{A Hybrid Approach to Temporal Pattern Matching} 

\author{\IEEEauthorblockN{Konstantinos Semertzidis}
\IEEEauthorblockA{\textit{IBM Research} \\
\textit{Dublin, Ireland}\\
konstantinos.semertzidis1@ibm.com}
\and
\IEEEauthorblockN{Evaggelia Pitoura}
\IEEEauthorblockA{\textit{Dept. of Computer Science \& Engineering} \\
\textit{University of Ioannina, Greece}\\
pitoura@cse.uoi.gr}}

\maketitle

\begin{abstract}
The primary objective of graph pattern matching is to find all appearances of an input graph pattern query in a large data graph. Such appearances are called matches.
In this paper, we are interested in finding matches of  interaction patterns in temporal graphs.
To this end, we propose a hybrid approach that achieves effective filtering of  potential
matches based both on structure and time. Our approach exploits  a graph representation where edges are ordered by time. 
We present experiments with real datasets that illustrate the
efficiency of our approach.
\end{abstract}

\section{Introduction}

\begin{figure}[ht!]
\centering
\resizebox{1.0\columnwidth}{!}{
\includegraphics[width=0.8\columnwidth]{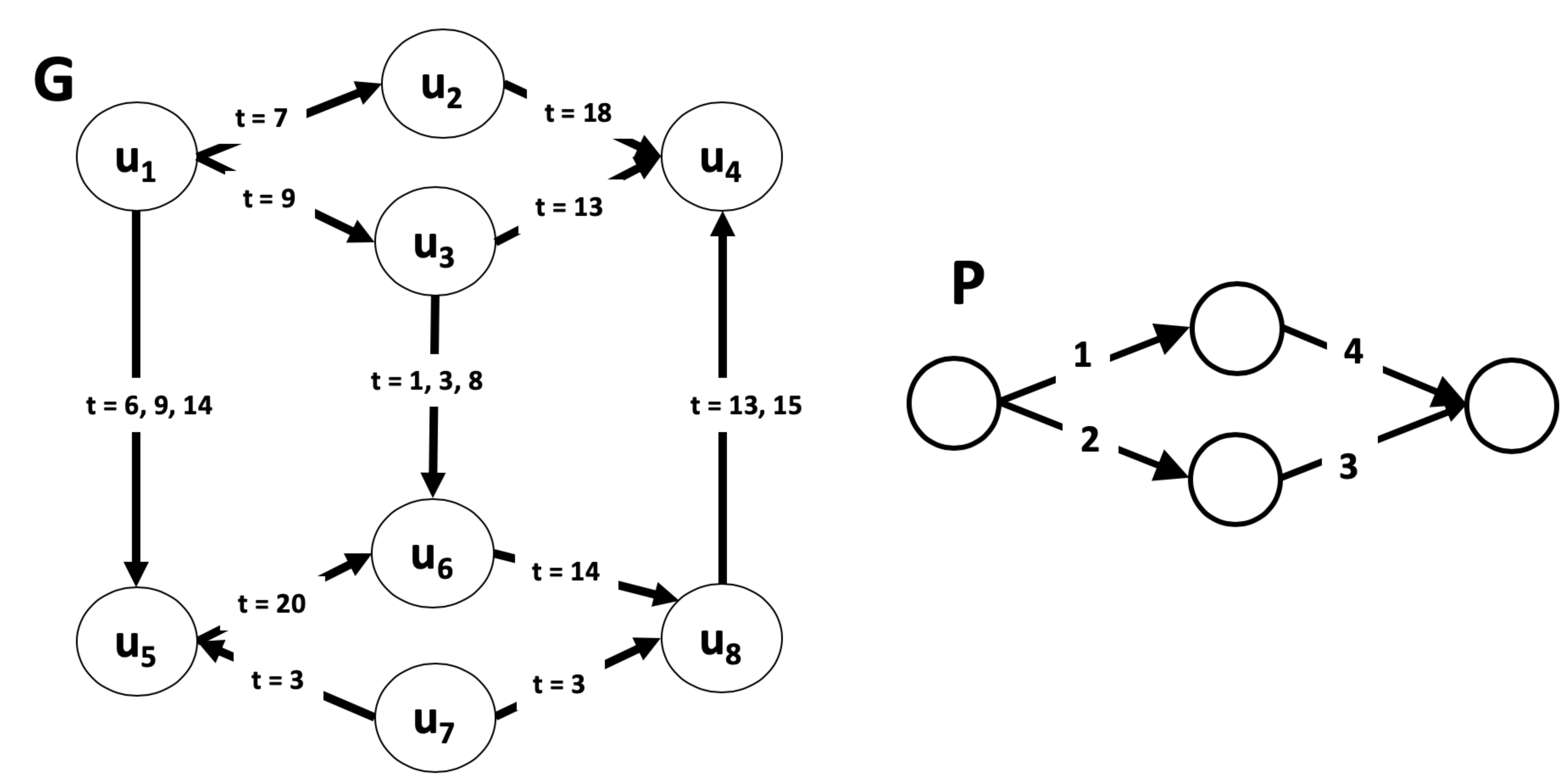}
}
\caption{Temporal graph $G$ and graph pattern $P$. For clarity, multiple edges between nodes in $G$ are shown as a single edge with  multiple timestamps.}
\label{fig:graph}
\end{figure}

In this paper, we focus on graphs whose edges model interactions between entities.
In such graphs, each edge is timestamped with the time when the corresponding interaction took place. 
For example, a phone call network can be represented as a sequence of timestamped 
edges, one for each phone call between two people.
Other examples include biological, social and financial transaction networks.
We are interested in finding patterns of interaction within such graphs.
Specifically, we assume that we are given as input a graph pattern query $P$ whose edges are ordered and this order specifies  the desired order of appearance of the corresponding interactions.
We want to find all matches of $P$ in a temporal graph $G$, that is, the subgraphs of $G$ that
match $P$ structurally, and whose edges respect the specified time order. We also ask that
all interactions in the matching subgraph appear within
a given time interval $\delta$. An example interaction pattern $P$ and temporal graph $G$ are shown in Fig. \ref{fig:graph}.

We propose an efficient algorithm which uses an edge-based representation of the graph where edges are ordered based on time.
This representation allows fast pruning of the candidate matches that do not meet the temporal constraints. 
We then extent this representation to achieve combined structural and temporal pruning. 
Our experiments on four real datasets show the efficiency of our approach.

\noindent\textbf{Related Work:} There has been recent interest in processing and mining  temporal graphs, including among others discovering communities \cite{DBLP:conf/icdm/ZhouCZG07}, computing measures such as density \cite{DBLP:conf/cikm/GalimbertiBBCG18, DBLP:journals/datamine/SemertzidisPTT19}, PageRank \cite{DBLP:conf/wise/HuZG15, DBLP:conf/pkdd/RozenshteinG16}, and shortest path distances \cite{DBLP:conf/ssdbm/HuoT14, DBLP:conf/adbis/SemertzidisP17,DBLP:journals/pvldb/WuCHKLX14}. 
There have also been previous work on locating temporal motifs and subgraphs \cite{DBLP:conf/sigmod/GurukarRR15, kovanen2011temporal, DBLP:journals/pvldb/KumarC18, DBLP:conf/wsdm/ParanjapeBL17,  DBLP:conf/asunam/RedmondC13, DBLP:journals/tkde/SemertzidisP19, DBLP:conf/edbt/ZufleREF18}.
The work in \cite{DBLP:journals/tkde/SemertzidisP19} introduces the problem of finding the most durable matches of an input graph pattern query, that is, the matches that exist for the longest period of time in an evolving graph.
An algorithm for finding temporal subgraphs is provided in \cite{kovanen2011temporal}, with the restriction that edges in the motif must represent consecutive events for the node.
Temporal cycles are studied in \cite{DBLP:journals/pvldb/KumarC18} and a heuristic approach for computing an estimation of the numbers of temporal motifs is studied in \cite{DBLP:conf/sigmod/GurukarRR15}. 
The authors of \cite{DBLP:conf/asunam/RedmondC13} consider a query graph with non time-stamped edges and define as temporal match a subgraph such that all its  incoming edges occur before its outgoing edges.

Most closely related to ours is the work in  \cite{DBLP:conf/wsdm/ParanjapeBL17}, and \cite{DBLP:conf/edbt/ZufleREF18}.
The authors of \cite{DBLP:conf/edbt/ZufleREF18} index basic graph structures in the static graph so as to find a candidate set of isomorphic subgraphs quickly at query
time, and then  verify for each candidate match whether the time conditions hold.
The authors of  \cite{DBLP:conf/wsdm/ParanjapeBL17} use as similar to ours time ordered representation of the temporal graph but they locate the temporal graphs by a static subgraph matching algorithm and then use a separate algorithm to find all temporal subgraph matches.
In summary, these works enumerate matches of a pattern graph using a two phase approach: they first perform structural matching and then they filter the results that do not meet the time conditions. 
Our work proposes a combined structural-temporal matching approach.

\section{Problem definition}
A \textit{temporal graph} $G(V, E)$ is a directed graph where  $V$ is a set of nodes, $E$ is a set of \textit{temporal edges} $(u, v, t)$ where $u$ and $v$ are nodes in $V$, and $t$ is a timestamp in $\mathbb{R}$.
There may be multiple temporal edges between a pair of nodes capturing interactions appearing at different time instants.
An example is shown in Fig. \ref{fig:graph}.

We use $\Pi(u, v)$ to denote the set of edges from $u$ to $v$, $\pi(u,v)$
the number of such edges, and $time(e)$ the timestamp of temporal edge $e$.  
For example in Fig. \ref{fig:graph}, the nodes $u_1, u_5$ are connected through three edges, e.g, $\Pi(u_1, u_5) = \{e_1, e_2, e_3\}$ with $time(e_i) = 6, time(e_2) = 9$, and $time(e_3) = 14$.

The same timestamp may appear more than once modeling simultaneous interactions. Thus, timestamps impose a partial order $\prec_{\textsc{t}}$ on temporal edges.
For a pair of temporal edges,  $e_i$ and  $e_j$ in $E$, $e_i$ $\prec_{\textsc{t}}$ $e_j$, if $time(e_i)$ $<$ $time(e_j)$ and $e_i$ $=_{\textsc{t}}$ $e_j$, if $time(e_i)$ $=$ $time(e_j)$.

We define the start time, $start\_time(G)$,  and  end time, $end\_time(G)$, of a temporal graph $G$ as  the smallest and largest timestamp appearing in any of the temporal edges in $E$. We also define the duration, $dur(G)$ of a temporal graph as  $end\_time(G)$ - $start\_time(G) + 1$.

We are interested in finding patterns of interactions in a temporal graph that appear within a time period of $\delta$ time units.
A $\delta$-interaction pattern $P$ is a temporal graph with $dur(P)$ $\leq$ $\delta$.

\begin{definition} \textit{[Interaction Pattern Matching]}	
Given a temporal graph 	$G(V, E)$, a time window $\delta$ and a $\delta$-interaction pattern $P$ = $G(V_P, E_P)$, a subgraph $M$ = $G(V_M, E_M)$ of $G$ is a match of $P$, such that the following conditions hold:
\begin{enumerate}
	\item (subgraph isomorphism) There exists a bijection $f$: $V_P \rightarrow V_M$ such that, there is a one-to-one mapping between the edges in $E_P$ and $E_M$, such that.  for each edge ($u$, $v$, $t$) in $E_P$, there is exactly one edge 
	($f(u)$, $f(v)$, $t'$) in $E_M$,
	\item (temporal order preservation) For each pair of edges $e_1$ and $e_2$ in $E_P$  mapped to edges $e'_1$ and $e'_2$  in $E_M$, it holds $e_1$ $\prec_{\textsc{t}}$ $e_2$ $\Rightarrow$ $e'_1$ $\prec_{\textsc{t}}$ $e'_2$ and
 $e_1$ 	$=_{\textsc{t}}$ $e_2$ $\Rightarrow$ $e'_1$ $=_{\textsc{t}}$ $e'_2$.
	\item (within $\delta$)  $dur(M)$ $\leq$ $\delta$.
\end{enumerate}
\end{definition}

For example, in Fig. \ref{fig:graph}, the subgraph $G'$ induced by $\{u_1,u_2,u_3,u_4\}$ is a match of $P$ with $dur(G') =  12$. 


\section{Algorithms}

We call \textit{static graph} the directed graph $G_S(V, E_S)$  induced from  temporal graph $G(V, E)$, if we ignore timestamps and multiple edges, i.e., $(u, v)$ is an edge in $E_S$, if and only if, there is a temporal edge  $(u, v, t)$ in $E$.

A straightforward approach is to apply a graph pattern matching algorithm on the static graph $G_S$ and then for each match find all time-ordered matches.  However, it is easy to see that
for a match $M_S$  in $G_S$,  the number of candidate time-ordered matches can be up to  $\prod_{i=1}^{|E_{M_S}|} \pi(e^{M_S}_{i})$, for each $e_i \in M_S$. In the following, we introduce an alternative approach.


\subsection{Interaction Pattern Matching} 
Our pattern matching algorithm identifies the interaction pattern matches by traversing a representation of the graph where edges are ordered based on $\prec_{\textsc{t}}$.
Let $L_G$ be this representation, that is, the list of graph edges of $G$ ordered by $\prec_{\textsc{t}}$.
We also order by $\prec_{\textsc{t}}$  the edges in the pattern, let $L_P$ = $e^P_1, \dots e^P_{|E_P|}$ be the resulting list.
We match pattern edges following the order specified in the input pattern.
The algorithm matches the  edges in the pattern edge-by-edge in a depth-first manner until all edges are mapped and a full matching subgraph is found.
The basic steps are outlined in Algorithm \ref{alg:edge-search}. 
We use a stack $S$ to maintain the graph edges that have been mapped to pattern edges so far.
Index $i$ denotes the pattern edge  to be mapped next.
{\sc MatchingEdge} performs two types of pruning:
\begin{enumerate} 
\item [(1)] temporal pruning: matched edges must (a) follow the time-order and (b) be within $\delta$. 
\item [(2)] structural pruning: the matched graph must be isomorphic to the pattern subgraph.
\end{enumerate}
Next, we describe a simple variation of {\sc MatchingEdge} termed 
{\sc SimpleME} and then a more efficient one termed {\sc IndexME}.

\begin{algorithm}[ht!]\small
	\caption{\small InteractionSearch($G$, $P$, $\delta$)}
	\label{alg:edge-search}
	\begin{algorithmic}[1]
		\Require{Temporal graph $L_G$, pattern graph $L_P$,  time window $\delta$}
		\Ensure{Interaction matches $m$}
		\vspace{0.1cm}
		\hrule
		\vspace{0.1cm}
		\State $S \leftarrow \emptyset$ \Comment{stack with candidate matches}
		\State $M \leftarrow \emptyset$ \Comment{interaction matches}
		\State $sT \leftarrow \delta$ \Comment{remaining time}
		\State $MoreMatches$ $\leftarrow$ $True$; $i$ $\leftarrow$ $1$
		\While{$(MoreMatches)$}
		\State $e^G_{i} \leftarrow$ {\sc MatchingEdge}$(e^P_{i}, sT)$
		
		\If {$e^G_i \neq null$}
		\State $Push( e^G_i, S)$
		\If {$i = |E_P|$} \Comment{a compete match is found}
		\State $M$ $\leftarrow$ $M$ $\cup$ $Content(S)$
		\State $i$ $\leftarrow$ 1
		\Else $~~i \leftarrow i + 1$; $sT \leftarrow sT - time(e^G_{i})$
		\EndIf
		\Else \Comment{no match found}
		\If {$i = 1$} $MoreMatches$ $\leftarrow$ $False$
		\Else $~~Pop(S)$;  $i \leftarrow i - 1$
		\EndIf	
		\EndIf
		\EndWhile
	\end{algorithmic}
\end{algorithm}

\spara{Simple Temporal-Structural filtering.}
{\sc SimpleME} shown in Algorithm \ref{alg:simple-find-match}
 scans the graph edges in  $L_G$ linearly. 
The key idea is that when we have matched the $i$-th pattern edge with a graph edge at position $j$, to match the next $i + 1$-th pattern edge, we need to look only at positions in  $L_G$ larger than   $j$.
As we build the match, we maintain the mapping $F$ of pattern edges to graph edges. We use $F(u)$ to denote the graph node that pattern node $u$ was mapped to. If $u$ is not yet mapped, $F(u)$ is $null$.

\begin{algorithm}[ht!]\small
	\caption{\small SimpleME($(u^P, v^P)$, $rT$)}
	\label{alg:simple-find-match}
	\begin{algorithmic}[1]
		\Require{Pattern edge $(u^P, v^P)$ to be mapped, remaining time $rT$, mapping $F$ of pattern nodes to graph nodes, position in the graph of the previously matched edge $i^G$ }
		\Ensure{Graph edge $(u^G, v^G)$  matching pattern edge $(u^P, v^P)$}
		\vspace{0.1cm}
		\hrule	
		\vspace{0.1cm}
		\For {($i=i^G + 1$ to  $max$)}
		\State $(u^G, v^G) = L_G[i]$ 
		\If {($time(u^G, v^G)$ $\leq$ $rT$)}
		\If {$(F(u^P) = F(u^P) = null$) \textbf{or}  $(F(u^P) = null$ \textbf{and} $F(v^P) = v^G)$ \textbf{or} $(F(v^P) = null$ \textbf{and} $F(u^P) = u^G$) \textbf{or} $(F(u^P) = u^G$ \textbf{and} $F(v^P) = v^G$) }
		\State $i^G = i$   \Comment{matching edge found}
		\State Update map $F$
		\State \Return{($(u^G, v^G)$)}
		\EndIf
		\Else
		 \State  $i$ $\leftarrow$ $i$ + 1
		\EndIf
		\EndFor
		\State	\Return(null) \Comment{no matching edge found}
	\end{algorithmic}
	\vspace{-0.1cm}
\end{algorithm}

Let $i_G$ denote the last position in  $L_G$ where  a graph edge has been mapped to a pattern edge.
In {\sc SimpleME}, we just go through the edges of the graph starting just after  the position of the previously matched edge (position $i^G$ + 1). For each edge, we check whether the edge satisfies the temporal and graph isomorphism conditions.
The maximum number of edges to be checked is equal to the number of edges $e_i$ for which
$time(e_i) \leq  rT + time(e_{i^G})$. For example, if $rT = 5$ and the time of the last  mapped edge is 4 then only the edges $e_i$ with $time(e_i) \leq 9$ should be checked.

\spara{Indexed temporal-structural filtering.}  
{\sc IndexME} uses information about the ingoing and outgoing edges of each node to avoid the linear scanning of the graph. Two  structures are used: (a)
an additional mapping structure $F_{edge}$, and (b) an extension of the graph structure $L_G$
with neighborhood information.

Specifically, in addition to the mapping table $F$, we maintain a mapping table $F_{edge}$.
Assume that pattern node $u^P$ during the course of the algorithm is mapped to graph node $u^G$.  We maintain the matching graph edge $u^G$ in $F(u^P)$, and also in $F_{edge}(u^P)$ the
most recently matched graph edge for which $u^G$ was either a source node or a target node. 
Note that this edge is also the most recent edge. All graph edges that will match the remaining pattern edges must appear later in
time than this edge.

We also extend $L_G$. For each edge $e^G$ = $(u^G, v^G)$ in $L_G$, we maintain in $N_{s}(u^G, e)$ the next in time edge with source node $u^G$, in $N_{t}(u^G, e)$
the next in time edge with target node $u^G$ and similarly
$N_{s}(v^G, e)$ and $N_{t}(v^G, e)$ for node $v^G$.  
We can retrieve all in and out neighbors of a node ordered by time by simply following these lists.
For example, to find the out-neighbors of node $u$, we start by the first edge, say $e$, where $u$ appears, retrieve $N_{s}(u, e)$, then $N_s(u, N_{s}(u, e))$  and so on.
We will use the notation $next_{in}(u)$ ($next_{out}(u))$ to denote getting the next in-neighbor 
(resp. out-neighbor) of $u$.
Note, that  each edge is represented by its position in the $L_G$ graph.

\begin{algorithm}[ht!]\small
	\caption{\small IndexME($(u^P, v^P)$, $rT$)}
	\label{alg:index-find-match}
	\begin{algorithmic}[1]
		\Require{Pattern edge $(u^P, v^P)$ to be mapped, remaining time $rT$, mapping $F$ of pattern nodes to graph nodes, position in the graph of the previously matched edge $i^G$, list $L_{in}$  and $L_{out}$}
		\Ensure{Graph edge  $(u^G, v^G)$  matching pattern edge $(u^P, v^P)$}
		\vspace{0.1cm}
		\hrule
		\vspace{0.1cm}		
	\If {$F(u^P) = null$ \textbf{and} $F(v^P) = null$}
	
\For {($i=i^G + 1$ to  $max$)}
		\State $(u^G, v^G) = L_G[i^G]$ 
			\If {($time(u^G, v^G)$ $\leq$ $rT$)}
		\State $i^G = i$   \Comment{matching edge found}
		\State Update $F(u^P)$, $F_{edge}(u^P)$, $F(v^P)$, $F_{edge}(v^P)$ 
			\State \Return{($(u^G, v^G)$)}
		\Else $~~i$ $\leftarrow$ $i$ + 1
		\EndIf
	\EndFor
	\State	\Return(null) \Comment{no matching edge found}
	
	\ElsIf {$F(u^P) \neq null$ \textbf{and} $F(v^P) \neq null$}
	\State  $j$ $\leftarrow$ $max \{F_{edge}(u^P), F_{edge}(v^P) \}$ 

		\For {($i=j$ to  $max$)}
	\If {$j$ = $F_{edge}(u^P)$}
	\State search the out-neighbors of $N_s(L_G[j]$ to find	$F(v^P)$ 
	\Else $~~$ search the in-neighbors of $N_t(L_G[j]$ to find	$F(u^P)$ 
	\EndIf
	\State Let $m$ be the edge found
	\If {($time(m)$ $\leq$ $rT$)} 
	\State $i^G = m$   \Comment{matching edge found}
 \State Update  $F_{edge}(u^P)$, $F_{edge}(v^P)$ 
	\State \Return{($(m)$)}
\Else $~~i$ $\leftarrow$ $i$ + 1
\EndIf
\EndFor
\State	\Return(null) \Comment{no matching edge found}
\ElsIf {$F(u^P) \neq null$}
\State search as above using next out $N_s(L_G[j])$ 
\Else $~~i$ search as above using next in $N_t(L_G[j])$ 
	\EndIf
	\end{algorithmic}
\end{algorithm}

Let us now describe the  {\sc IndexME} algorithm (shown in Algorithm \ref{alg:index-find-match})  in detail. The algorithm first checks whether any of the pattern edge endpoints has already been mapped.
If none of them has already been matched, then in Lines 2-9, it searches for any match satisfying the
time constraints in Line 4.
Otherwise, if both endpoints have been matched in Lines 10-22, we get one of the endpoints, let this be node $u$ and depending on whether it is a  source or a target node, either the $next_{out}(u)$ or $next_{in}(u)$ is used to find the edges containing the other endpoint in Lines 13-15.
Next, if the mapped edge is active within the remaining time window the algorithm updates the matched edge index structure and returns the edge in Lines 17-20.
In case just one  of the two endpoints is matched, say $u$, we follow the same steps for missing endpoint $v$ by looking into the in-neighbor (resp. out-neighbor) lists of $u$ in Lines 23-25.

\section{Experimental Evaluation}
\label{sec:exp}

We use the following real datasets \cite{DBLP:conf/wsdm/ParanjapeBL17}:
(1) \textbf{Email-EU} contains emails between members of a European research institution; an edge indicates an email send from a person $u$ to a person $v$ at time $t$.
(2) \textbf{FBWall} records the wall posts between users on Facebook located in the New Orleans region; each edge indicates the post of a user $u$ to a user $v$ at a time $t$.
(3) \textbf{Bitcoin} refers to the decentralized digital currency and payment system. It consists of all payments made up to October 19, 2014; an edge records the transfer of a bitcoin from address $u$ to address $v$ at a time $t$.
(4) \textbf{Superuser} records the interactions on the stack exchange web site Super User;
edges indicate the users' answers/comments/replies to comments.
The characteristics of the datasets are summarized in Table \ref{table:properties}.


\begin{table}[h!]
\centering
\caption{Dataset characteristics}
\resizebox{1.0\columnwidth}{!}{
	\begin{tabular}{c c c c c}
		\hline
		\textbf{Dataset} & \textbf{\# Nodes} & \textbf{\# Static Edges} & \textbf{Edges} &\textbf{\# Time (days)}\\
		\hline
		Email-EU & 986 & 24,929 & 332,334 & 803\\
		FBWall & 46,952 & 274,086 & 876,993 & 1,506 \\
		Bitcoin & 1,704 & 4,845 & 5,121,024 & 1,089\\
		Superuser & 194,085 & 924,886 & 1,443,339 & 2,645\\
		\hline
	\end{tabular}}
	\label{table:properties}
\end{table}

\begin{table}[ht!]
\caption{Execution time (sec) of $IPM$ and $Competitor$ for varying time window $\delta$ and path lengths.}
\resizebox{1.0\columnwidth}{!}{
\centering
\begin{tabular}{@{}|c||c|c|c|c|c|c|c|c|@{}}
\toprule
    & \multicolumn{2}{c|}{\small {Email-EU}}       & \multicolumn{2}{c|}{\small{ FBWall}}    & \multicolumn{2}{c|}{\small {Bitcoin}}      & \multicolumn{2}{c|}{\small {Superuser}}       \\ \midrule\hline
 $\delta$ (hours)  & IPM  & Comp & IPM  & Comp & IPM & Comp & IPM          & Comp \\ \midrule \hline
$\delta=1$ & 1.45 &  21  &  3.72 &  42 &  1.48 &  15.7 & 4.07  &  45\\ \midrule
$\delta=3$ &  1.56 & 49 &  3.85 &  65 &  4.24 & 135  & 4.92 & 66 \\ \midrule
$\delta=6$&   1.86 & 115 &  3.92  &  105  & 15.7 & 675 &  4.94 &  101 \\ \midrule
$\delta=9$ &   2.85 & 171 &  4.17 &  132  & 48.8  &  $>$ 0.5h &  5.3 & 127\\ \midrule
$\delta=12$ &  3.41 & 190 &  4.65 &  152  & 113  &  $>$ 0.5h &  5.5 & 160\\ \midrule  
\hline
 Path length & IPM  & Comp & IPM  & Comp & IPM & Comp & IPM & Comp \\ \midrule \hline
$l=2$ & 0.05 &  0.3 &  3.4 & 10 & 1 &   6.5  & 3.35 & 15  \\ \midrule
$l=4$ &  0.9 &  1.05 &  3.5 &  28 &  1.1 & 10 &  3.8 & 18   \\ \midrule
$l=6$ &  1.45 &  21 &  3.72  & 42 & 1.48  &  15.7 &  4.07 &  45\\ \midrule
$l=8$ & 1.9 &  103 &  4 &  80 &  1.63 &  28 & 4.93 & 110   \\ \midrule
$l=10$ & 2.2 &  115 &  4.28 &  85 &  1.72 &  36 & 5.53 & 121   \\ \bottomrule
\end{tabular}}
\vspace{-0.3cm}
\label{table:baseline}
\end{table}

\spara{Results:}
We first compare  our algorithm, termed $\ed$, with the competitive approach followed in \cite{DBLP:conf/wsdm/ParanjapeBL17, DBLP:conf/asunam/RedmondC13} which first generates the candidate subgraphs that match the topology of the pattern and then filters the results that do not follow the specified temporal order. 
The generation of candidate subgraphs is done edge by edge and only the matches
active within the $\delta$ time window are retained.
As our default pattern queries, we use  two type of queries:
(a) path queries, and (b) random graph pattern queries.
Path pattern queries represented as a sequence of edges with consequent timestamps.
Random graph pattern queries are generated as follows. For
a random query of size $n$, we select a node randomly from the
graph and starting from this node, we perform a DFS traversal
until the required number n of nodes is visited. We use as our pattern, the graph created by the union of visited nodes and traveled edges. We set the ordering of edges using the topological ordering of the graph.
We report the average performance of 100 random queries for each size n.

Table \ref{table:baseline} reports the results a) for various $\delta$ (for $\delta = 1$ up to $12$ hours) and path length = 6, and b) for various path lengths (from 2 up to 10) and fixed $\delta = 1$ hour. 
As shown, the $\baseNode$'s response time  increases with $\delta$, while $\ed$ is considerably faster in all datasets.
This is because the $\baseNode$ tends to generate many redundant matches which do not follow the required temporal order. 
This is more prominent when we increase the query duration $\delta$.
Also, there are cases where a found match $m$ contains edges with multiple timestamps and thus the algorithm must generate a large number of temporal matches. This is the case especially with the Bitcoin  and the Email-EU datasets, where there are multiple timestamped edges per static edge.
Regarding the path length, the $\baseNode$'s response time is increasing with the path length while, again, $\ed$ is considerably faster and  not affected much by the query size. 
$\ed$ takes advantage of the indexes and verifies for each edge with a few lookups whether  there is any other valid edge within the remaining time window. Similar results hold for random queries.
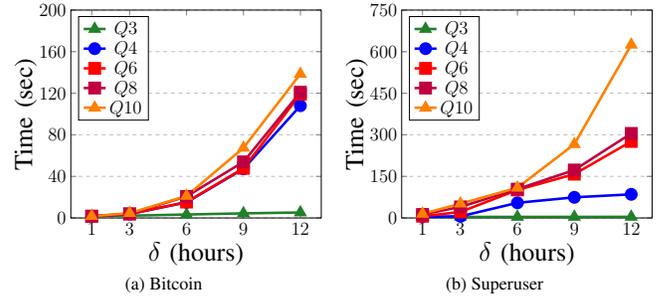
\begin{figure}[ht!]
\centering
\resizebox{1.0\columnwidth}{!}{
\begin{tabular}{cccc}
\subfloat[Bitcoin]
{\pgfplotsset{scaled y ticks=false}
\begin{tikzpicture}
\begin{axis}[
	xtick = {1, 3, 6, 9, 12},
	legend style={legend columns=1, font=\Large, anchor=north,at={(0.5,1.15)}},
	xticklabel style = {font=\Large},
	yticklabel style = {font=\Large},
	xlabel={$\delta$ (hours)},
	ylabel={Time (sec)},
	ytick distance=40,
	ymax=200,
	tick label style={/pgf/number format/fixed},
	y label style={font=\huge},
    x label style={font=\huge},	
    ymin=0,
    legend pos=north west,
	ymajorgrids=true,
	grid style=dashed,
]
	\addplot[color=darkgreen, mark=triangle*,line width=2pt,mark size=4pt]coordinates{(1,1.5)(3,2.18)(6,3.298)(9,4.425)(12,5.265)};
	\addplot[color=blue, mark=*,line width=2pt,mark size=4pt]coordinates{(1,1.6)(3,3.803)(6,15.01)(9,47.228)(12,108)};
	\addplot[color=red, mark=square*,line width=2pt,mark size=4pt]coordinates{(1,1.681)(3,3.826)(6,15.298)(9,47.78)(12,119)};
\addplot[color=purple, mark=square*,line width=2pt,mark size=4pt]coordinates{(1,1.688)(3,3.963)(6,20.739)(9,53.95)(12,121)};
\addplot[color=orange, mark=triangle*,line width=2pt,mark size=4pt]coordinates{(1,1.719)(3,4.645)(6,21.16)(9,67.449)(12,138.517)};
\legend{$Q3$, $Q4$, $Q6$, $Q8$, $Q10$}
\end{axis}
\end{tikzpicture}} &
\subfloat[Superuser]
{\pgfplotsset{scaled y ticks=false}
\begin{tikzpicture}
\begin{axis}[
	xtick = {1, 3, 6, 9, 12},
	xticklabel style = {font=\Large},
	yticklabel style = {font=\Large},
		legend style={legend columns=1, font=\Large, anchor=north,at={(0.5,1.15)}},
	xlabel={$\delta$ (hours)},
	ylabel={Time (sec)},
	ytick distance=150,
	ymax=750,
	tick label style={/pgf/number format/fixed},
	y label style={font=\huge},
    x label style={font=\huge},	
    ymin=0,
    legend pos=north west,
	ymajorgrids=true,
	grid style=dashed,
]
	\addplot[color=darkgreen, mark=triangle*,line width=2pt,mark size=4pt]coordinates{(1,3.225)(3,3.693)(6,3.787)(9,3.9)(12,4.025)};
	\addplot[color=blue, mark=*,line width=2pt,mark size=4pt]coordinates{(1,4.22)(3,5.848)(6,54.6)(9,74.446)(12,85.25)};
	\addplot[color=red, mark=square*,line width=2pt,mark size=4pt]coordinates{(1,6.3)(3,21.63)(6,101)(9,158)(12,276)};
\addplot[color=purple, mark=square*,line width=2pt,mark size=4pt]coordinates{(1,12.71)(3,39.95)(6,104)(9,173)(12,305)};
\addplot[color=orange, mark=triangle*,line width=2pt,mark size=4pt]coordinates{(1,15)(3,51.61)(6,109)(9,266)(12,625)};
\legend{$Q3$, $Q4$, $Q6$, $Q8$, $Q10$}
\end{axis}
\end{tikzpicture}}
\end{tabular}}
\caption{{Query time for random queries for various query sizes.}}
\label{fig:random}
\end{figure}

Fig. \ref{fig:random} reports the performance of our algorithm for random queries of different graph sizes $Qn$ (from $n=2$ up to $n=10$ nodes) and different time windows.
All small queries are processed very fast because the algorithm requires only a few lookups to identify adjacent nodes within the given time window. Moreover, even for largest queries and longest timer windows where larger lookups are required, our algorithm only takes
$\sim$2 and $\sim$10 minutes to process the query in Bitcoin and
Superuser respectively.  The performance difference in Superuser
dataset is expected since it consists  of a much larger number of
edges and time instants than Bitcoin.

\section{Conclusions}
In this paper, we studied the problem of locating matches of patterns of interactions in temporal graphs that appear within a specified time period.
We  presented an efficient algorithm based on a representation of graph where edges are ordered based on their interaction time.
Our experimental evaluation with real datasets demonstrated the efficiency of our algorithm in finding time-ordered matches.
In the future, we plan to study 
the  streaming version of the problem where given a stream of graph updates we locate the interaction patterns occurring within a sliding time window. 

\bibliographystyle{plain}
\bibliography{bibliography} 
\end{document}